\documentclass[twocolumn,showpacs,preprintnumbers,amsmath,amssymb]{revtex4}
\usepackage{graphicx}% Include figure files

\begin{document}

\title{Influence of Rb, Cs and Ba on Superconductivity of Magnesium Diboride.}

\author{A.~V.~Palnichenko}
 \email{paln@issp.ac.ru}
\author{O.~M.~Vyaselev}
\author{N.~S.~Sidorov}
\affiliation{Institute of Solid State Physics, Chernogolovka,  Moscow district., 142432 Russia}

\date{\today}

\begin{abstract}

Magnesium diboride has been thermally treated in the presence of Rb, Cs, and Ba. Magnetic susceptibility shows onsets of
superconductivity in the resulting samples at 52~K (Rb), 58~K (Cs) and 45~K (Ba). Room-temperature $^{11}$B NMR indicates to
cubic symmetry of the electric field gradient at boron site for the samples reacted with Rb and Cs, in contrast to the axial
symmetry in the initial MgB$_2$ and in the sample treated with Ba.

\end{abstract}

\pacs{74.25.Ha, 74.25.Nf, 74.62.Bf, 74.70.Ad, 74.70.Dd}
\maketitle

Recent discovery of superconductivity in magnesium diboride, MgB$_2$, with $T_c \simeq$ 39~K \cite{Nagam} has stimulated
intensive study of this material. Its simple chemical composition, high symmetry of crystal lattice ($P6/mmm$) and hence high
symmetric electronic structure \cite{Kortus} have enabled fine calculations to clarify the mechanism of high temperature
superconductivity in MgB$_2$ \cite{Mazin}.

It has been reported previously that doping MgB$_2$ with carbon \cite{Takenoby}, Li \cite{Zhao, Cimberle}, Be \cite{Felner}, Zn
\cite{Kazak}, Al \cite {Cimberle, Slusky}, Ti \cite{Zhao1}, Ni, Fe, Co \cite {Mori}, Mn \cite{Mori1}, and Si \cite{Cimberle}
only reduces the superconducting transition temperature. No information has been reported so far about doping of MgB$_2$ with
heavier alkali and alkali-earth metals. These elements are capable of strong carrier donation to the electron system and may
therefore essentially enhance superconducting properties of the host material. In this paper we report on the influence of Rb,
Cs and Ba on superconductivity of MgB$_2$.

MgB$_2$ was thermally treated with Rb, Cs and Ba through liquid-phase reaction. The reacted samples are hereafter nicknamed as
$\&$Rb, $\&$Cs, and $\&$Ba samples. Two types of initial material were used: One contained 0.2-1~mm size granules of sintered
in 1:1 molar ratio mixture of MgB$_2$ and Mg (MgB$_2$/Mg). The other one was fine powder of pure MgB$_2$. The initial material
was vacuum sealed with the abundance of the metal (Rb, Cs or Ba) in quartz ampules preliminarily evacuated down to
10$^{-3}$~Torr. The reactions were held with Rb and Cs for 10-100~hours at 160-300$^\circ$C, and with Ba for 5 minutes at
700$^\circ$C.

Superconductivity of the samples was studied by measuring the magnetic susceptibility, $\chi$, in temperature interval
4.2-300~K. The low-frequency ac susceptibility was measured using a home-made set-up by means of a mutual inductance technique
at frequency 623~Hz in a driving field $\sim$0.1~Oe. The signal corresponding to the real part of $\chi$ was detected using a
lock-in amplifier. Superconducting transitions of led, niobium, and BiSCO:2212 ($T_c$=90~K) were used to insure fidelity of the
temperature scale of the measurements, as well as for calibration of the diamagnetic response. In the case of Rb the
measurements were made without unsealing the ampules to avoid sample oxidation. It also enabled to consecutively alter heat
treatment and measurement on the same batch. The unreacted excess of the metal heat-linked the particles of the sample and the
walls of the vacuumized ampule to provide correct measurement of the sample temperature.

\begin{figure}[ht]
\includegraphics[width=1\linewidth]{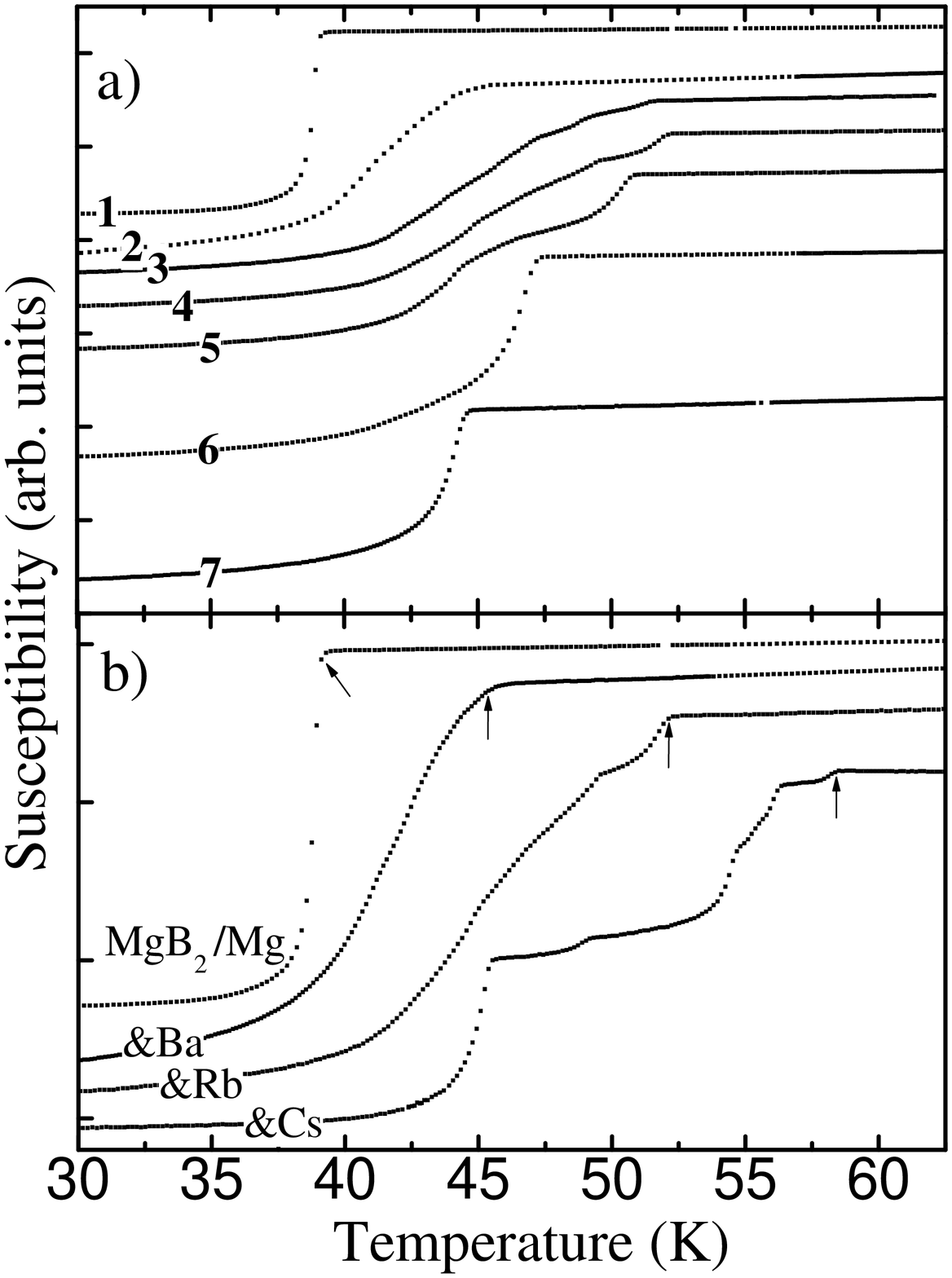}
\caption{\label{Suscept} Temperature dependencies of magnetic susceptibility: (a) measured between consecutive heat treatments
of MgB$_2$/Mg with Rb, 1 -- initial sample, 2 -- 10~hours at 180$^\circ$C, 3 -- 13~hours at 190$^\circ$C, 4 -- 15.5~hours at
200$^\circ$C, 5 -- 36~hours at 200$^\circ$C, 6 -- 56~hours at 200$^\circ$C, 7 -- 18~hours at 300$^\circ$C; (b) the initial
MgB$_2$/Mg, the one reacted with Ba (5~min. at 700$^\circ$C), with Rb (17~hours at 200$^\circ$C), and with Cs (20~hours at
160$^\circ$C followed by 100~hours annealing at 100$^\circ$C). Arrows indicate onsets of superconducting transitions.}
\end{figure}

To track the changes in the superconducting properties during the treatment of MgB$_2$ with an alkali metal, the reaction with
Rb was performed through portions of heat treatment alternated by susceptibility measurements. Each time two samples were
simultaneously exposed to the heat treatment, one containing MgB$_2$/Mg and the other one with pure MgB$_2$. Temperature
dependencies of magnetic susceptibility, $\chi$($T$), measured between consecutive portions of heat treatment on MgB$_2$/Mg
with Rb, are shown in Figure~\ref{Suscept}a. $\chi$($T$) of the host MgB$_2$/Mg sample before the heat treatment, depicted in
curve 1, has a sharp drop at 39~K denoting the superconducting transition at this temperature. Both the onset temperature and
the width of the transition coincide with those of pure MgB$_2$\cite{Nagam}. Subsequent heat treatments with Rb lead to first
increase of the transition onset temperature up to 52~K (curves 2 through 4), and then rollback to 44~K (curves 5 through
7)\cite{WhySC}. The sample that initially contained pure MgB$_2$ powder has retained the superconducting transition at
$\simeq$39~K during the whole sequence of heat treatment. Hence under the applied conditions of thermal treatment, admixture of
Mg to MgB$_2$ plays the key role in formation of the superconducting phase with $T_c$ higher than in the host MgB$_2$.

Relying on the above results, the heat treatment regimes  were adjusted for the reactions of MgB$_2$/Mg with Cs and Ba.
Fig.~\ref{Suscept}b shows $\chi$($T$) plots for  $\&$Cs and $\&$Ba samples, as well as $\chi$($T$) for the initial MgB$_2$/Mg
and for the $\&$Rb sample. The $\chi$($T$) curves for $\&$Cs and $\&$Ba samples in Fig.~\ref{Suscept}b deviate towards the
diamagnetic state below 58~K and 45~K, respectively, indicating the onsets of superconductivity. Therefore the superconducting
transition temperature of the host MgB$_2$/Mg can be essentially enhanced through the reaction with Rb, Cs, or Ba.

\begin{figure}[b]
\includegraphics[width=1\linewidth]{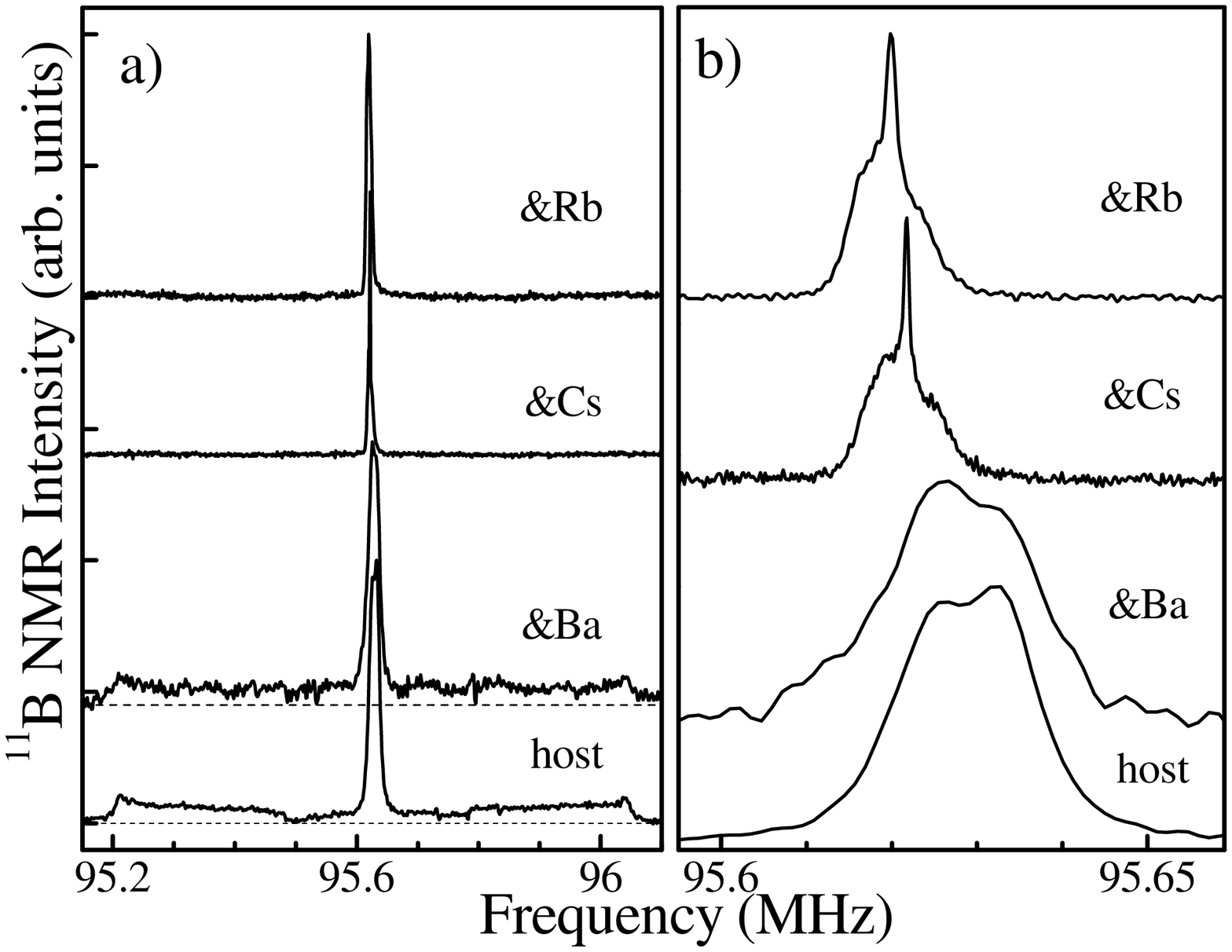}
\caption{\label{NMR}  Room-temperature $^{11}$B NMR spectra at 7~T in the host MgB$_2$/Mg and in MgB$_2$/Mg reacted with Ba,
Cs, and Rb: The spectra in whole (a) and details of the central peaks (b).}
\end{figure}

The samples characterized in Fig.~\ref{Suscept}b were subjected to $^{11}$B NMR. The spectra were taken at room temperature in
7~T field using a Bruker MSL-300 spectrometer by means of the standard spin-echo technique with $\pi$ pulse of $\simeq4~\mu$s.
Since $^{11}$B NMR spectrum is known to be $\sim$800~kHz wide \cite{VerkhovMgB2,BorsaMgAlB2} and the experiment bandwidth
limited by the $\pi$-pulse length is $\sim$100kHz, for each sample several spectra between 95 and 96.2~MHz at frequencies
spaced by 100~kHz have been collected and added up. The resulting spectra are shown in Fig.~\ref{NMR}.

It is known from previous studies \cite{VerkhovMgB2,BorsaMgAlB2} that $^{11}$B NMR spectrum in pure MgB$_2$ includes a sharp
central peak and two broader satellite peaks separated by $\simeq$830~kHz. This kind of NMR spectrum  originates from the spin
3/2 of $^{11}$B and the axial symmetry of the charge environment of boron in MgB$_2$. The central peak corresponding to
($-1/2\leftrightarrow1/2$) transition is close to the boron Larmour frequency,  $^{11}\nu_L$=$^{11}\gamma H_0$, where
$^{11}\gamma $ is $^{11}$B gyromagnetic ratio and $H_0$ the external magnetic field. The satellites peaks
($\pm3/2\leftrightarrow\pm1/2$ transitions) result from the quadrupole interaction of the nucleus with the charge environment
in the crystal. At high (compared to quadrupole interaction) fields the distance between the satellites equals to the
quadrupole frequency, $\nu_Q$, proportional to the principal value of the electric field gradient (EFG) at the nucleus site,
$V_{zz}$ .

$^{11}$B NMR spectra in MgB$_2$/Mg and in MgB$_2$/Mg reacted with Ba (Fig.~\ref{NMR}a) agree with the results reported
previously for pure MgB$_2$ \cite{VerkhovMgB2,BorsaMgAlB2} including the same $\nu_Q\simeq$830~kHz separating the satellites
and the Pake doublet-topped central peak, Fig.~\ref{NMR}b, with $\simeq$17~kHz full width at half maximum, FWHM. This infers
that Ba essentially changes neither the axial symmetry nor the value of EFG at boron site. On the contrary, the spectra in the
$\&$Rb and $\&$Cs samples differ strongly from that of the host MgB$_2$/Mg. First, the satellite intensities are much weaker
and practically invisible in Fig.~\ref{NMR}a implying the presence of a phase with cubic symmetry of EFG at boron site with
$V_{zz}$=0, besides the native axially symmetric phase. Comparison of the satellites and the central peak areas in MgB$_2$/Mg
and in the $\&$Rb and $\&$Cs samples yields about 30$\%$ of the axially symmetric phase in the samples with alkali metals.
Secondly, the central peaks in the $\&$Rb and $\&$Cs samples (Fig.~\ref{NMR}b) consist of an extremely narrow
(FWHM$\simeq1$~kHz) feature on top of a wider (FWHM$\simeq7$~kHz) pedestal and are lower in frequency by $\simeq$8~kHz=84~ppm.
Since the cental $^{11}$B NMR peak in MgB$_2$ is broadened by the second-order quadrupole interaction \cite{BorsaMgAlB2}, the
smaller linewidth of the central peaks in the samples with alkali metals also indicates to cubic symmetry of the EFG at boron
site. More detailed analysis of the NMR data will be reported in a separate publication.

In the summary, the superconducting properties of MgB$_2$ reacted with Rb and Cs at 160-300$^\circ$C, and with Ba at
700$^\circ$C have been studied. The compounds having superconducting transitions higher than in MgB$_2$ have been found for the
samples that initially contained 1:1 mixture of MgB$_2$ and Mg. Room-temperature $^{11}$B NMR has indicated to the presence of
a phase with cubic symmetry of charge environment at boron site in the samples reacted with Rb and Cs, in contrast to the
axially symmetric cases for the host MgB$_2$ and for the sample reacted with Ba.

We thank Dr. O.~G.~Rybchenko for structural characterization of the samples, prof. V.~V.~Ryazanov and prof. Yu.~A.~Ossipyan for
fruitful discussions. The work has been partially supported by the Russian Foundation for Basic Research, Grant No~05-02-17731.

\end{document}